\documentclass[aps,prl,multicol,reprint,preprintnumbers, footbib]{revtex4-2} 
\usepackage{float}
\usepackage{balance}
\newcommand{\squeezeup}{\vspace{-2.5mm}}
\newcommand{\squeezeupl}{\vspace{-1.8mm}}

\usepackage{graphics}
\usepackage{bm}        
\usepackage{amssymb}   
\usepackage{xcolor}
\usepackage{graphicx}
\usepackage{subfigure}
\usepackage[colorlinks=true,urlcolor=blue,linkcolor=blue]{hyperref}

\usepackage{array}
\usepackage{mathtools,slashed}
\usepackage{multirow}
\usepackage{rotating}

\newcommand{\uL}{\underline{\mL}}

\newcommand{\gR}{g_{_{R}}}

\newcommand{\mB}{\mathsf{B}}
\newcommand{\mL}{\mathsf{L}}
\newcommand{\bF}{\boldsymbol{W}}

\newcommand{\bW}{\boldsymbol{W}}

\newcommand{\p}{\partial}

\newcommand{\bPhi}{\boldsymbol{\Phi}}

\newcommand{\an}{\quad \textmd{and} \quad }
\newcommand{\bs}{\boldsymbol{\sigma}}

\newcommand{\bea}{\begin{eqnarray}}
\newcommand{\eea}{\end{eqnarray}}

\newcommand{\bN}{{\bf{N}}}
\newcommand{\bn}{{\boldsymbol{\nu}}}

\newcommand{\where}{\quad \textmd{where} \quad}
\newcommand{\x}{\tilde{\tau}}

\newcommand{\bk}{{\bf{k}}}

\newcommand{\bD}{{\boldsymbol{\Delta}}}

\newcommand{\mpl}{M_{\mbox{\tiny{Pl}}}}

\begin{document}

\preprint{CERN-TH-2020-217}

\title{SU(2)$_R$ and its Axion in Cosmology: A common Origin for Inflation, Cold Sterile Neutrinos, and Baryogenesis}
\date{\today}
\author{Azadeh Maleknejad}
\affiliation{Theoretical Physics Department, CERN, 1211 Geneva 23, Switzerland
}
\email{azadeh.maleknejad@cern.ch}


\begin{abstract}
{We introduce an axion-inflation model embedded in the Left-Right symmetric extension of the SM in which $W_R$ is coupled to the axion. This model merges three milestones of modern cosmology, i.e., inflation, cold dark matter, and baryon asymmetry. Thus, it can naturally explain the observed coincidences among cosmological parameters, i.e., $\eta_{\mB}\approx P_{\zeta}$ and $\Omega_{DM} \simeq 5~\Omega_{\mB}$. The source of asymmetry is spontaneous CP violation in the physics of inflation, and the lightest right-handed neutrino is the cold dark matter candidate with mass $m_{N_1}\sim 1~GeV$.  The introduced mechanism does not rely on the largeness of the unconstrained CP-violating phases in the neutrino sector nor fine-tuned masses for the heaviest right-handed neutrinos. It has two unknown fundamental scales, i.e. scale of inflation $\Lambda_{\rm inf}=\sqrt{H\mpl}$ and left-right symmetry breaking $\Lambda_{F}$. Sufficient matter asymmetry demands $\Lambda_{\rm inf}\approx\Lambda_{F}$. The baryon asymmetry and dark matter today are remnants of a pure quantum effect (chiral anomaly) in inflation, which, thanks to flavor effects, are memorized by cosmic evolution.}

\end{abstract}

\maketitle
\flushbottom

The two pillars of the post-inflationary scenarios of leptogenesis are: i) CP asymmetric decay of massive Right-handed neutrinos (RHN) after reheating, and ii) washout processes to enhance the efficiency and eliminate the pre-existing asymmetry to avoid theoretical uncertainties \cite{Kuzmin:1985mm}. Besides, the lightest sterile neutrino may account for the Dark Matter (DM) \cite{Dodelson:1993je}. The source of CP asymmetry is the CP-violating phases in the neutrino sector, unconstrained by the current data, which are assumed to be large enough. 
Moreover, low-scale leptogenesis mostly requires fine-tuning of parameters, e.g. highly degenerate RHN masses, e.g. $m_{N_3}\simeq m_{N_2}$ \cite{Canetti:2012kh}.
Besides, once flavor effects are considered, it is difficult for a pre-existing asymmetry to be washed out by the RH neutrino decays \cite{Bertuzzo:2010et} (
See also Fig. \ref{fig:flavor}).

This \textit{letter} introduces a new framework for simultaneous baryogenesis and dark matter production within General Relativity (GR), which avoids the above issues. The source of asymmetry is spontaneous CP violation by a $W_R$ gauge field coupled to the inflaton that produces leptons and baryons in inflation. In this scenario, baryon asymmetry and DM are remnants of the same effect in inflation. Thus it can naturally explain the observed coincidences among cosmological parameters, i.e., $\eta_{\mB}\approx P_{\zeta}$ and $\Omega_{DM}\simeq5\Omega_{\mB}$. 

Early universe physics is a subject that seeks
answers for fundamental questions linking the very high energy physics (immensely small scales) with the extremely large cosmological scales. The Standard Model (SM) of particle physics, highly successful in formulating fundamental particles at low energy scales, is greatly incomplete when it meets cosmology and astrophysics. The most glaring shortcomings of the SM are (I) the neutrino mass, (II) baryon asymmetry of the Universe (BAU), and (III) particle nature of dark matter. Considering cosmic inflation as the leading paradigm for early universe \cite{Guth:1980zm}, we should, as well, add (IV) the particle nature of the inflaton field to this list. SM as a theory for particle physics also faces a number of issues, i.e., (i) Higgs vacuum stability problem,  (ii) accidental $\mB-\mL$ global symmetry, and (iii) \textit{ad hoc} parity violation at the Electro-Weak scale (EW). 


\begin{figure}[h!]
   \centering \includegraphics[width=\columnwidth,height=0.25\textheight]{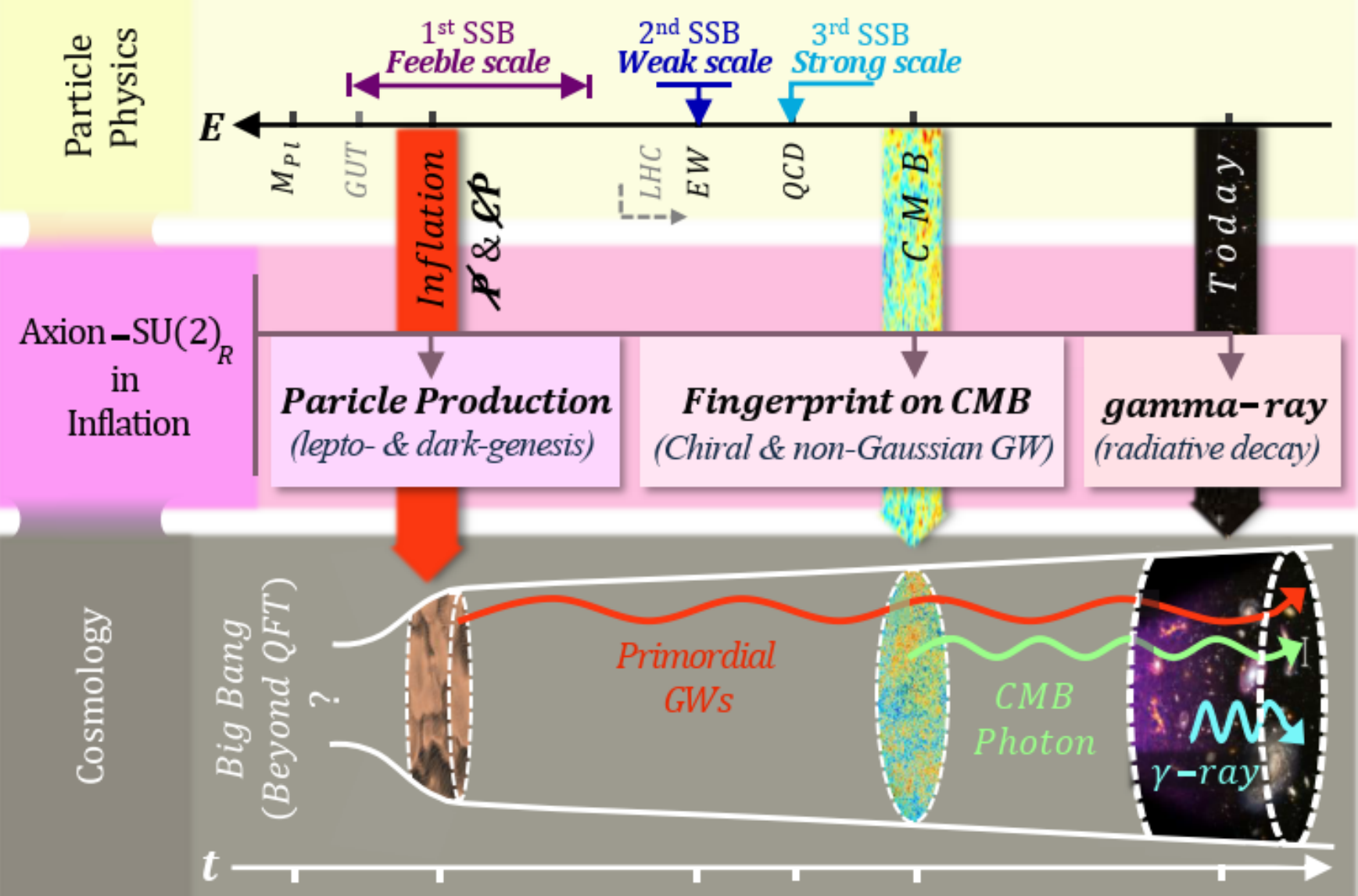}\\
\squeezeup
  \caption{
  $SU(2)_R$-axion inflation: a natural common origin for inflation, fermionic dark matter, and matter asymmetry and its observational signatures. }
  \label{fig:Cosimic-History} 
\end{figure}
\squeezeup


Axion fields are well-motivated candidates for inflaton and are naturally coupled to gauge fields. As first discovered by the author, non-Abelian gauge fields may survive inflation and contribute to its physics while respecting the cosmological symmetries  \cite{Maleknejad:2011sqq}. That introduced a new class of inflation models accompanied by $SU(2)$ gauge fields with an immensely rich phenomenology. The minimal realization of this idea is an $SU(2)$ gauge field in GR coupled to a generic axion inflaton \cite{Maleknejad:2011sqq, Adshead:2012kpp, Maleknejad:2016qjzz}, so-called $SU(2)$-axion inflation.  For a review on gauge fields in inflation, see \cite{Maleknejad:2012fww}. The novel features shared by these models are 
(1) Spontaneous $P$ and $CP$ violation, and satisfying all Sakharov conditions  in inflation \cite{Maleknejad:2016dcii}, (2)  Particle production by the gauge field in inflation through Schwinger effect \cite{Lozanov:2018kpkk, Maleknejad:2018nxzz, Maleknejad:2019hdrr} and chiral anomaly \cite{Maleknejad:2019hdrr}, (3) Naturally warm inflation \cite{Kamali:2019ppii}, (4) Prediction of chiral and non-Gaussian gravitational wave (GW) background \cite{Maleknejad:2012fww, Adshead:2013qpp, Agrawal:2017awzz} detectable by future CMB missions and laser interferometers \cite{Campeti:2020xwnn}. Therefore, this inflation setup merges three open issues of the SM and cosmology, i.e., inflation, DM, and BAU. It gains an additional value due to its unique observable signature on GW background induced by GW-$SU(2)$ field interactions. The connection of this $SU(2)$ field with the SM, however, was still missing, and in the present work, we attempt to fill the gap.

The aim of this \textit{letter} is to embed the $SU(2)$-axion inflation setting in gauge extensions of the SM and study its phenomenological and cosmological consequences. The most well-motivated Beyond the SM (BSM) theories are SUSY, GUT, and Left-Right Symmetric Models (LRSM) of the weak interactions. We restrict the current work to the most minimal realization, i.e., LRSM. 
Originally proposed to explain $P$ violation in low energy processes \cite{Pati:1974yy}, LRSM predicted massive neutrinos years before experiment. Among its additional appealing features are: natural $\mB-\mL$ symmetry \cite{Mohapatra:1980qee}, entailed seesaw mechanisms \cite{Mohapatra:1980ypp}, and solution to vacuum stability problem at high scales \cite{Maiezza:2016ybz}.

In this \textit{letter}, we assume that the gauge field in the $SU(2)$-axion inflation models is $W_R$ in the LRSM. Comparing with the minimal LRSM, here we have an axion $\varphi$, which is coupled to the $SU(2)_R$ and drives cosmic inflation. We call this particle physics model for inflation $SU(2)_R${\it{-axion inflation}}. This model is a complete setup that can simultaneously provide plausible explanations for the phenomena (I-IV) and (i-iii) named earlier. A more detailed analysis is presented in a followup work \cite{Maleknejad:2021nqi}.
The present paper can be a starting point for further, more involved analysis of the rich and multifaceted phenomenology of these gauge extensions of the SM in inflation physics.

\vspace{0.41mm}

\begin{center}{\bf{SU(2)$_R$-Axion Inflation Model}}
\end{center}
\vspace{-0.7mm}

The minimal gauge group that implements the hypothesis of left-right symmetry is $\mathcal{G} \equiv SU(2)_L\times SU(2)_R \times U(1)_{\mB-\mL}$ (suppressing color). The theory includes three gauge fields $W_{L,R}$ and $A_{\mB-\mL}$ associated with
$SU(2)_{L,R}$ and $U(1)_{\mB-\mL}$ respectively. The fermionic content is the SM quarks and leptons extended by three RHNs as
\bea
q_{iL,R} = ~\begin{pmatrix}
{\boldsymbol{u}}_{i} \\ {\boldsymbol{d}}_{i}
\end{pmatrix}_{\!L,R} \an 
l_{iL,R} = ~\begin{pmatrix}
{\bn}_{i} \\ {\boldsymbol{l}}_{i}
\end{pmatrix}_{\!L,R},
\eea
where $\bn_{iR}$ are three RHNs interacting by $SU(2)_R$. It is accompanied by an extended Higgs sector consists of a Higgs bi-doublet $\bPhi$, and $SU(2)_{L,R}$ triplets $\bD_{L,R}$. The Spontaneous Symmetry Breaking (SSB) structure of the LRSM is 
\bea
\mathcal{G} \xrightarrow[\text{1st SSB}]{T<\Lambda_F}  SU(2)_L \times U(1)_{Y} \xrightarrow[\text{2nd SSS}]{T<\Lambda_{W}} U(1)_{em}.\nonumber
\eea
Below the scale $\Lambda_F$, the first SSB happens which breaks the LR symmetry and gives a VEV to the $SU(2)_R$ triplet, i.e. $\langle \bD_R \rangle \neq 0$. At this point, $W^{\pm}_R$, $Z_R$, and $\bN_i\equiv \bn_i+\bn_i^c$ become massive. Next, when the temperature gets below EW scale, $T<\Lambda_{W}$, the Higgs bi-doublet acquires a VEV, i.e. $\langle \bPhi \rangle \neq 0$, and second SSB occurs which gives Dirac mass to the SM particles, active neutrinos included \cite{Mohapatra:1980ypp}. 

Now we add the inflaton, i.e. axion field, which is coupled to the $\bW_R$. As a concrete example we consider 
\bea
S_{\rm Inf} &=& \int d^4x \sqrt{-g} \big[  -\frac12 (\p_{\mu} \varphi)^2 - V(\varphi) + \mathcal{L}_{_{W_R}}\big],\\
\mathcal{L}_{_{W_R}} &=& - \frac12 {\rm{Tr}}[\bF_{\mu\nu}\bF^{\mu\nu}]_R - \frac{\lambda\varphi}{f} {\rm{Tr}}[\bF_{\mu\nu}\tilde{\bF}^{\mu\nu}]_R,
\eea
where $\bF_{R\mu\nu}$ is the strength tensor of $\bF_R$, $\tilde\bF^{\mu\nu}_R\equiv\frac12\frac{\epsilon^{\mu\nu\lambda\sigma}}{\sqrt{-g}} \bF_{R\lambda\sigma}$, $f\lesssim 10^{-1} \mpl$, and $\lambda \lesssim 1$. For the sake of generality, we assume $V (\varphi)$ is an arbitrary axion potential, flat enough to support the slow-roll inflation. For instance an axion monodromy inspired potential form \cite{McAllister:2014mpa}. This $SU(2)$-axion inflation model and its cosmic perturbations has been studied in \cite{Maleknejad:2016qjzz}.
The $SU(2)_R$-axion inflation has two unknown fundamental scales, i.e., the scale of inflation $\Lambda_{\rm inf}=\sqrt{\mpl H}$, and LR symmetry breaking $\Lambda_F$. Moreover, $\bF_R$ may or may not have a VEV in inflation. Thus, we can distinguish four different types of scenarios.
\squeezeupl
\begin{table}[h!]
\begin{center}
\begin{tabular}
{|c|c|c|}
\hline
 & \qquad $\Lambda_{\rm inf} > \Lambda_{F}$ \qquad \quad & \qquad $\Lambda_{\rm inf} < \Lambda_{F}$ \qquad \quad \\
 \hline
\quad $\langle \bF_{R} \rangle =0$ \qquad \quad & \qquad {\bf{I}}~ \qquad \quad & \qquad {\bf{II}}~ \qquad \\ 
\hline
\quad $\langle \bF_{R} \rangle \neq0$ \qquad \quad & \qquad ${\bf{I_{v}}}$ \qquad \quad & \qquad ${\bf{II}_{v}}$ \qquad \\ 
\hline
$\begin{matrix} \textmd{Mass in}  \\ \textmd{Inflation} \end{matrix}$ & $\begin{matrix} m_{W_R} =0 \\ ~m_{N_i}=0 \end{matrix}$ & $\begin{matrix} m_{W_R} \neq 0 \\ ~m_{N_i}\neq 0 \end{matrix}$  \\ 
\hline
\end{tabular}
\end{center} 
\end{table}
\vspace{-7mm}

Scenarios $\rm{\bf{I}}$ and $\rm{\bf{I}}_{v}$ describe the case $\Lambda_{\rm inf}>\Lambda_{F}$, while $\rm{\bf{II}}$ and $\rm{\bf{II}}_{v}$ when $\Lambda_{\rm inf}<\Lambda_{F}$. The $v$ subscript denotes systems in which the $SU(2)_R$ acquires a VEV in inflation. In this work, we focus on scenarios $\rm{\bf{I}}$ and $\rm{\bf{II}}$, $\langle \bW_R \rangle =0$, and leave $\rm{\bf{I}}_v$ and $\rm{\bf{II}}_v$ cases for future works. The RH fermions are coupled to the $\bW_R$ field and its axion as 
\bea
\mathcal{L}_{\Psi_{JR}} \supset  \bar{\Psi}_{JR} \big( \bs^{\mu} [iD_{\mu} + \gR \bW_{R\mu }] - \frac{\tilde\lambda \dot\varphi}{f}\big) \Psi_{JR},
\eea
where $\tilde\lambda$ is a constant, $D_{\mu}$ is the spinor covariant derivative, and RH fermions are collectively shown as
\bea
\Psi_{JR} = \{q_{iR},l_{iR}\} \quad \textmd{where} \quad (J=1,\dots, 6).
\eea

\begin{center}{\bf\bf{Particle Production in Inflation}}
\end{center}
\vspace{-1mm}

Due to conformal symmetry, the gauge fields associated with $SU(3)_c \times SU(2)_L \times U(1)_{\mB-\mL}$ group, as well as all the left-handed fermions, are exponentially decaying in inflation. However, the $\bW_R$ associated with $SU(2)_R$ is coupled to the inflaton and sourced by it. Subsequently, the generated $SU(2)_R$ gauge field produces RH fermions. Note that the axion cannot create Weyl fermions \cite{Weinberg:1996kr} and they are merely produced by $W_R$.
The main particle physical consequences of this setup as the inflation physics are: {\it{i)}} $P$ and $C$ are maximally broken by the chiral nature of the $SU(2)_R$ interaction with the axion, {\it{ii)}} $CP$, $\mB$, and $\mL$ are all violated by the non-perturbative effects of the $\bW_{R}$, i.e. chiral (Adler-Bell-Jackiw) anomaly \cite{Adler:1969gkk}, {\it{iii)}} $\mB-\mL$ is conserved (violated) in scenario type {\bf{I}} (type {\bf{II}}), while $\mB-\mL_{_{\rm SM}}$ is violated in both scenarios, and {\it{iv)}} out of thermal equilibrium condition holds during inflation. Thus, all the Sakharov conditions required for a BAU \cite{Sakharov:1967djj} are satisfied in inflation. The field equation of $\bW_{R}$ is 
\bea\label{gauge-field-eq}
(\p_{\mu}-i\gR \bF\!_{R\mu})\big[\bF^{\mu\nu}_R+\frac{\lambda\varphi}{f} \tilde{\bF}^{\mu\nu}_R\big]-m_{W_R}^2 \bF^{\nu}_R=0,~~
\eea
where $\gR$ is the gauge coupling of $\bW_R$. A massless gauge field (type {\bf{I}}) with momentum $\bk$ has two (transverse) polarization states specified by the polarization vectors $e^{\pm}(\bk)$ where $\bk.e^{\pm}(\bk)=0$. The massive gauge field (type { \bf{II}}) has an extra (longitudinal) mode with polarization vector $e^{3}(\bk)=\bk/k$ and its zero component coupled to it given by the constraint equation. Interestingly, the longitudinal mode and the zero component are decoupled from the axion and decay in inflation. Thus we neglect them and refer the interested reader to \cite{Maleknejad:2021nqi} for the detailed calculations. Now we define the following slowly increasing parameters
\bea
\xi \equiv \frac{\lambda \dot\varphi}{2fH} \an \tilde\xi \equiv \frac{\tilde\lambda}{\lambda} \xi.
\eea
Imposing the Bunch-Davies vacuum, the transverse modes $f^a_{~\pm}$, associated with $e^{\pm}(\bk)$ polarization states, are
\bea\label{fpm-a}
f^a_{~\pm}(\bk,\tau) = \frac{e^{i\kappa_{\pm}\pi/2}}{(2\pi)^{\frac32}\sqrt{2k}} W_{\kappa_{\pm},\mu}(2ik\tau),
\eea
where $\tau$ is conformal time, $W_{\kappa_{\pm},\mu}$ is the $W$-Whittaker function, 
$\kappa_{\pm} = \mp i\xi$, and $\mu^2 = \frac14 - \frac{m_{W_R}^2}{H^2}$.
The energy density of $W_R$ is
\bea\label{cT}
\langle \rho_{W_R} \rangle \simeq \bigg(\frac{H}{\mpl}\bigg)^2 \bar{\rho} ~\mathcal{T}(\xi,m_{W_R}),
\eea
where $\bar{\rho}=3\mpl^2H^2$, and $\mathcal{T}(\xi,m_{W_R})$ is a function of $\xi $ and $m_{W_R}$ shown in Eq. \eqref{cT-a}. For $\xi>m_{W_R}$, $\mathcal{T}$ increases (decreases) exponentially with the increase of $\xi$ ($m_{W_R}$) as
$\mathcal{T}(\xi,m_{W_R})\propto \frac{1}{(2\pi)^2} e^{2(\xi-\lvert \mu \rvert)  \pi}$, while for $\xi<m_{W_R}$, it has power-low behavior and softly increases with the increase of $m_{W_R}$ (See Fig. \ref{fig:cT}). During slow-roll, $\xi$ is an almost constant (gradually increasing) parameter. As a result, the axion slowly injects more and more energy into the gauge field sector, and inflation is warm.

The generated gauge boson field produces RH leptons and baryons in inflation.
The anomaly of baryon and lepton currents are respectively as
\bea\label{anomaly}
\nabla_{\mu} J^{\mu R}_{\mB} &=&  - \frac{3\gR^2 }{16\pi^2} {\rm{Tr}}[\bF^{\mu\nu}\tilde{\bF}_{\mu\nu}]_R,\\
\nabla_{\mu} J^{\mu R}_{\mL} &=&  - \frac{3\gR^2 }{16\pi^2} {\rm{Tr}}[\bF^{\mu\nu}\tilde{\bF}_{\mu\nu}]_R + 2im_{N_i} \bar{\bn}_{iR} \bn_{iR}.~~
\eea
However, the $\mB$ and $\mL$ violating interactions of the left-handed fermions remains negligible in inflation. The total lepton number is related to the SM one as
\bea
n_{\mL} = n_{\uL} + \sum_{i} n_{N_i}, \quad (\uL \equiv \mL_{SM})
\eea
where $n_{N_i} $ are the sterile neutrino lepton numbers. Using Eq. \eqref{fpm-a} in \eqref{anomaly}, we find the baryon and lepton numbers respectively as
\bea \label{nB}
n_{\mB} &\simeq & -\gR^2 H^3 \mathcal{K}(\xi,m_{W_R}),\\
\label{nL}
n_{\mL} &\simeq & - H^3 \big[ \gR^2 \mathcal{K}(\xi,m_{W_R}\!) + \sum_{i } \frac{\tilde\xi}{\pi} \big(\frac{m_{N_i}}{H}\big)^2 \mathcal{D}(\tilde\xi,m_{N_i}\!) \big],~~~~
\eea
where $\mathcal{K}(\xi,m_{W_R})$ is the contribution of chiral anomaly (a pure quantum effect) and $\mathcal{D}(\tilde\xi,m_{N_i})$ is the contribution of the mass term of RHNs (in type {\bf{II}} scenarios).  The explicit forms of these prefactors are presented in Eq.s \eqref{cK--} and \eqref{Int-n-m}, and their plots are shown in Fig. \ref{fig:cD}. The prefactor $\mathcal{K}(\xi,m_{W_R})$ increases (decreases) with the increase of $\xi$ ($m_{W_R}$) and for $\xi>M_{W_R}$ as
\bea
\mathcal{K}(\xi,m_{W_R}) \propto \frac{1}{(2\pi)^4} ~e^{2 \xi\pi }.
\eea
The prefactor $\mathcal{D}(\tilde\xi,m_{N_i})$ is of order one, and symmetric wrt $\tilde\xi$.
Net $\mB-\uL$ asymmetry ($\uL\equiv \mL_{SM}$) created by inflation is
\bea\label{alpha-def}
&& n^{\rm inf}_{\mB-\uL} = \sum_{i} n^{\rm inf}_{N_i} \simeq - \alpha_{\mB-\uL}^{\rm inf} H^3, \\  
&&\alpha_{\mB-\uL}^{\rm inf} \equiv \big[  \frac{\gR^2}{2} \mathcal{K}(\xi,m_{W_R}) +  \sum_{i } \frac{\tilde\xi}{\pi} \big(\frac{m_{N_i}}{H}\big)^2 \mathcal{D}(\tilde\xi,m_{N_i}\!)\big].\nonumber
\eea

\begin{center}{\bf{Evolution after Reheating}}
\end{center}
\vspace{-0.7mm}

The study of the post-inflationary evolution requires to specify our parameter space further. For the sake of concreteness, we restrict the current analysis by assuming the following conditions: C1) A hierarchical mass spectrum for the RH neutrinos (as implied by the neutrino oscillations) as
\bea\label{mass-h-RHN}
m_{N_3}\gtrsim 10^{12}~GeV \gg m_{N_2} \gtrsim 10^{9}~GeV \gg m_{N_1},
\eea
where $\bN_1$ is much lighter than the EW scale, and with feeble Yukawa interactions, i.e., a DM candidate.
C2) The CP-violating phases in the neutrino sector, unconstrained by the current data, are not enough to create the observed BAU. 
C3) The post-inflationary generation of RHNs with $W_R$ interactions via freeze-out and freeze-in is negligible compared to their pre-existing relics.

{\it{Memory effect and remnant asymmetries:}} The spectator effects have important impacts in the final values of $\mB$, $\uL$ and $\mL_{N_1}$ (DM relic density). These effects are discussed in Sec. \ref{Spectators}. The final $\mB$ and $\uL$ ($\equiv \mL_{SM}$) are
\bea
n_{\mB}(a) = 0.12 ~ n^{\rm inf}_{\rm B-\uL} ~\big(\frac{a_{\rm inf}}{a}\big)^3 \an
n_{\uL}(a) =  -\frac74 n_{\mB}(a) . \nonumber
\eea
$\bN_{3,2}$ decay to lighter particles, while $\bN_1$ freezes out soon after inflation. Due to its feeble Yukawa interactions it can account for the DM with a relic number density as
\bea
n_{N_1}(a) &=& - \frac{\gR^2H^3}{6} \mathcal{K}(\xi,m_{W_R}) ~\big(\frac{a_{\rm inf}}{a}\big)^3.
\eea
\vspace{-6mm}
\begin{figure}[h!]
  \centering\includegraphics[width=\columnwidth,height=0.07\textheight]{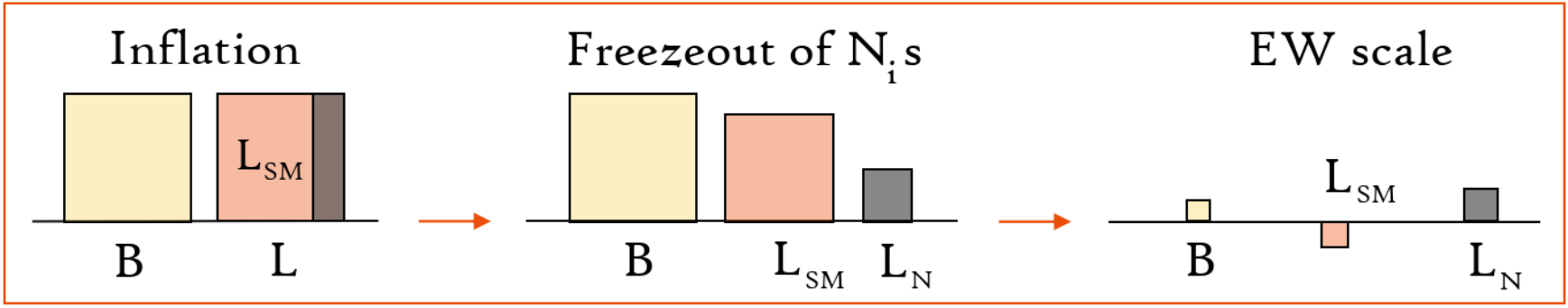}
  \vspace{-5mm}
   \caption{Evolution of $\mB$, $\uL$, and $\bN_1$ at three stages; (Left) Inflation, (Middle) Freeze-out of $N_i$s, and (Right) EW scale. }
  \label{fig:BSM} 
  \vspace{-3mm}
\end{figure}

{\it{Photon number density:}} Reheating starts at some point after the end of inflation. Here, we consider the phenomenological reheating model $\rho_{\rm reh}=\varepsilon \big(\frac{a_{\rm inf}}{a_{\rm reh}}\big)^{\!4}\rho_{\rm inf}$ in which $\varepsilon$ is the efficiency of the reheating process and relates $\rho_{\rm reh}$ and the energy density at the end of inflation, $\rho_{\rm inf}$ (See \ref{reheat}).
Thus the photon number density today ($g_{\rm eff,0}=\frac{43}{11}$) is 
\bea\label{photon}
n_{\gamma,0}= \frac{6\sqrt{3}\zeta(3)}{\pi^2 }  \big(\frac{\varepsilon}{g_{\rm eff,0}}\big)^{\frac34} (H\mpl)^{\frac32} \big(\frac{a_{\rm inf}}{a_0}\big)^3.
\eea
Note that, due to condition C3, entropy injection by the decay of heavier RHNs after reheating is negligible.

{\it{Demands imposed by C3:}}  Negligible freeze-out and freeze-in production of RHN by $W_R$ interactions requires 
 \bea\label{C3}
\frac{H}{\mpl} \lesssim \mathcal{A}\times 10^{-2}\varepsilon^{-\frac12} \big(\frac{a_{\rm reh}}{a_{\rm inf}}\big)^2 \bigg(\frac{1}{\gR} \frac{m_{W_R}}{\mpl}\bigg)^{\frac83},~~~
\eea
where $\mathcal{A}=\frac12\big( \frac{g_{\rm eff}}{10^2}\big)^{\frac12}$ is of the order one. C3 imposes an upper bound on the scale of inflation. For the details of the freeze-in production in this setup see \cite{Maleknejad:2021nqi}.

\vspace{1mm}
{\bf{Baryon to Photon Ratio:}} today we have
\bea\label{Eta-1}
\eta_{\mB,0} \equiv \frac{n_{\mB,0}}{n_{\gamma,0}} \simeq  -\frac{1}{3} \frac{\alpha_{\mB-\uL}^{\rm inf}}{\varepsilon^{\frac34}} \big(\frac{H}{\mpl}\big)^{\frac32}.
\eea
It is directly related to the amplitude of the primordial curvature power spectrum $P_{\zeta}(k)= P_{\zeta}(k_0)(\frac{k}{k_0})^{\!^{n_s-1}}$ as
\bea
\eta_{\mB,0} \simeq -\frac{2(2\pi)^2}{3} \bigg( \frac{\epsilon~\alpha_{\mB-\uL}^{\rm inf}}{\varepsilon^{\frac34}} \bigg) \big(\frac{\mpl}{H}\big)^{\frac12}~ P_{\zeta}(k_0),
\eea
where $\epsilon$ is the slow-roll parameter, $\eta_{B}\simeq 6\times 10^{-10}$, and $ P_{\zeta}(k_0)\approx 2 \times\! 10^{-9}$ \cite{Aghanim:2018eyx}.
The scale of inflation, then, is
\bea\label{etaB}
\frac{H}{\mpl} \simeq  10^{-6} \frac{\varepsilon^{\frac12}}{ \big(\alpha^{\rm inf}_{\mB-\uL}\big)^{\frac23}}.
\eea
Combining Eq.s \eqref{C3} and \eqref{etaB}, we find the allowed parameter space. Interestingly, it demands $\Lambda_F\approx \Lambda_{\rm inf}$, i.e. the LR SSB should coincide with the geometrical transition that ends inflation. More precisely, we need $H\mpl \approx \gR^{-2}m^2_{W_R}$. Moreover, the values of these parameters
are within the natural range of parameters in GUT theories \cite{Maleknejad:2021nqi}. For instance with a $\xi\in (2,4)$ and $\Delta N\equiv \ln(\frac{a_{\rm reh}}{a_{\rm inf}}) \gtrsim 2$, we find $\frac{1}{\gR} m_{W_R}\sim 10^{-4} ~\mpl$ and $H\sim 10^{-8}~\mpl$ (See Sec. \ref{reheat}).

\vspace{1mm}
{\bf{Cold Dark Matter Relic Density:}} The number density of $\bN_1$ neutrino today is $
n_{N_1,0} \simeq 2.8 ~ n_{\mB,0}$.
If it makes all the DM today, its mass is
\bea
m_{N_1} \approx 1.8 ~ m_{P} \simeq 1.7 ~GeV,
\eea
where $m_{P}$ is the proton mass. Since the production mechanism is independent of the active-sterile mixing angles, $\bN_1$ can have a lifetime much larger than the age of the Universe. Nevertheless, via its loop-mediated radiative decay channel, it can decay to gamma-ray photons with energy $E_{\gamma}\approx m_{N_1}/2$ \footnote{Demanding that $\bN_1$ is stable over the lifetime of the universe gives 
$\theta_1 < 10^{-13}$.
In this framework, the generation mechanism of $\bN_1$ is independent of its Yukawa mixing
with active neutrinos, and $\theta_1$ can be any number that satisfies the above upper bound.}. Thus, it may provide observable effects to be probed by gamma-ray telescopes.

\begin{center}{\bf\bf{Discussion}}
\end{center}
\vspace{-1mm}

This letter introduced the $SU(2)_R$-axion inflation model embedded in the Left-Right symmetric extension of the SM. It is a new framework for simultaneous baryogenesis and darkgenesis. 
 One of the most well-studied leptogenesis scenarios with new gauge interactions is the LRSMs \cite{Hati:2018tge,Dunsky:2020dhn}. Let us explore the differences between previous studies and the current proposal. The scenarios so far discussed in the literature rely on unconstrained CP-violating phases in the neutrino sector. {\textbf{(a)}} a relic abundance of RHNs is generated after reheating by $W_R$ interactions via freeze-out or freeze-in mechanisms. {\textbf{(b)}} the asymmetric decay of RHNs, then, creates matter asymmetry.  As an alternative mechanism, the current proposal set parameters such that phenomena {\textbf{(a)}}-{\textbf{(b)}} named earlier are negligible. 
The source of asymmetry is spontaneous CP violation in the physics of inflation, and the lightest right-handed neutrino is the cold dark matter candidate. Relic abundances of SM leptons, baryons, and RHNs are generated by the chiral anomaly of $W_R$ in inflation. 
Sufficient asymmetry does not require fine-tuned masses for the heaviest right-handed neutrinos, but it demands $\Lambda_{\rm inf}\approx\Lambda_{F}$. Therefore this new framework relates the scale of $SU(2)_R\times U(1)_{\mB-\mL}$ breaking with the end of inflation and prefers scales above $10^{10}~GeV$. Interestingly, it is in the range suggested by the non-supersymmetric SO(10) GUT with an intermediate left-right symmetry scale.
Due to the common origin of inflation, cold dark matter, and baryon asymmetry it can naturally explain the observed coincidences among cosmological parameters, i.e., $\eta_{\mB}\approx P_{\zeta}$ and $\Omega_{DM} \simeq 5~\Omega_{\mB}$ with $m_{N_1}\sim 1~GeV$.



{\it{Acknowledgments:~}} The author would like to thank Eiichiro Komatsu for insightful discussions and valuable input during previous collaborations on which part of the present work is based. She thanks Marco Drewes for helpful discussions. (In memory of Navid Afkari \& Ruhollah Zam)


%


\pagebreak
\widetext 

\begin{center}
\textbf{\large Supplemental Materials: \\
SU(2)$_R$ and its Axion in Cosmology: A common Origin for Inflation, Cold Sterile Neutrino, and Baryogenesis}
\vskip 0.3cm
Azadeh Maleknejad\\
\textit{Theoretical Physics Department, CERN, 1211 Geneva 23, Switzerland
}
\end{center}

\setcounter{secnumdepth}{2}
\setcounter{equation}{0}
\setcounter{figure}{0}
\setcounter{table}{0}
\setcounter{page}{1}
\setcounter{section}{0}
\makeatletter
\renewcommand{\theequation}{S\arabic{equation}}
\renewcommand{\thefigure}{S\arabic{figure}}
\renewcommand{\thetable}{S\arabic{table}}
\renewcommand{\thesection}{S\arabic{section}}


Here we merely discuss some essential calculations.  A more detailed analysis is presented in \cite{Maleknejad:2021nqi}. 

\section{Particle Production in Inflation}

\begin{figure}[htb]
\includegraphics[height=0.2\textheight]{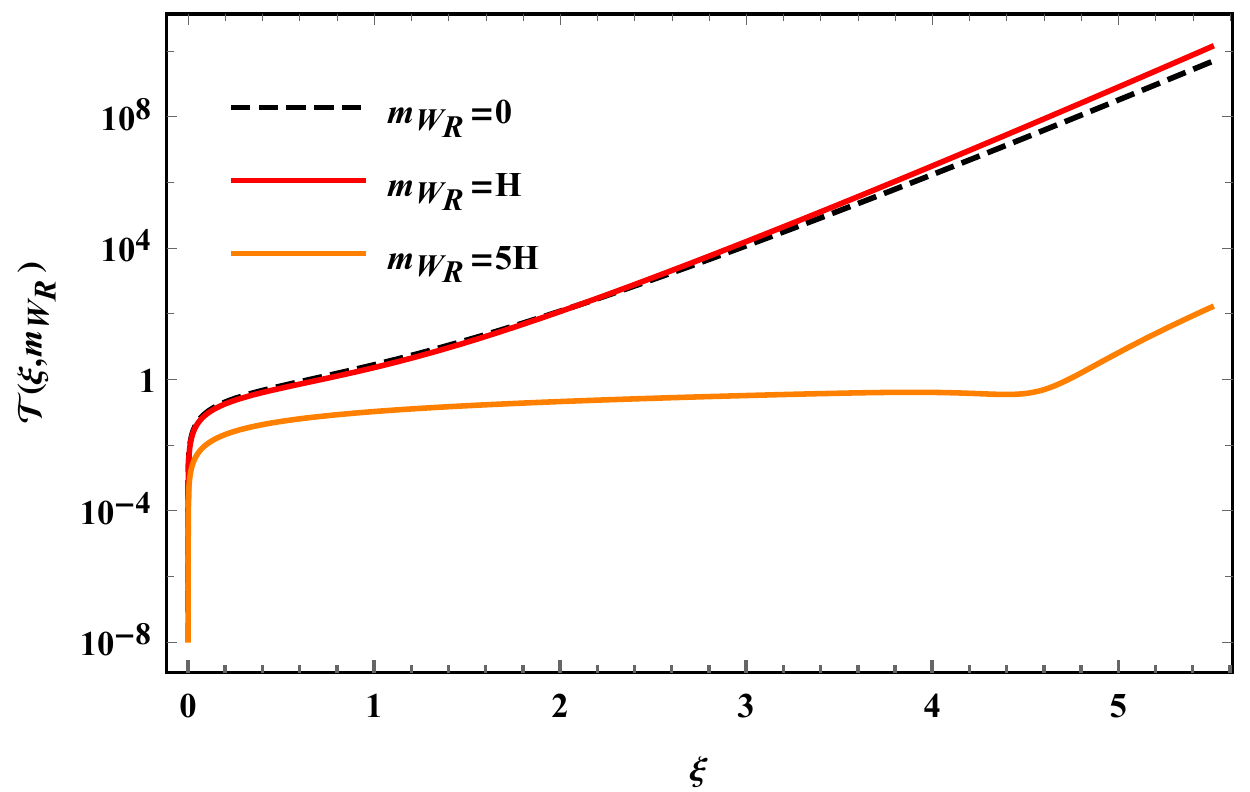}
\includegraphics[height=0.2\textheight]{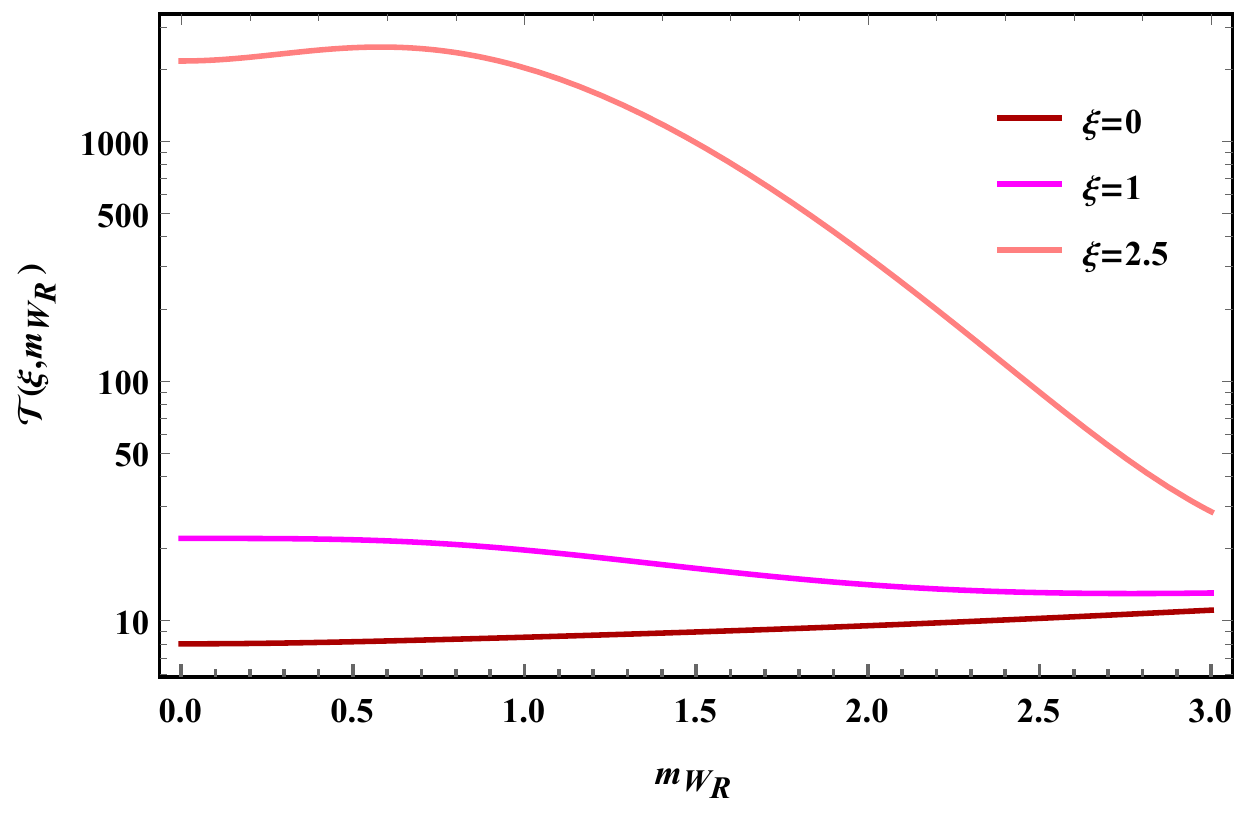}
\caption{The prefactor $\mathcal{T}(\xi,m_{W_R})$ in Eq. \eqref{cT} with respect to $\xi$ and $m_{W_R}$. The left panel shows $\mathcal{T}(\xi,m_{W_R})$ vs $\xi$ and the right panel shows it vs $m_{W_R}$. }
\label{fig:cT} 
\end{figure}

The prefactor  $\mathcal{T}(\xi,m_{W_R})$ in Eq. \eqref{cT} is
\bea\label{cT-a}
\mathcal{T}(\xi,m_{W_R}) \equiv  \sum_{\sigma=\pm} \frac{e^{\sigma \xi \pi}}{2(2\pi)^2} \int \x^4d\ln\x \bigg[  \p_{\x} W^{*}_{\kappa_\sigma,\mu}\p_{\x} W_{\kappa_\sigma,\mu} + (1 + \frac{m_{W_R}^2}{H^2\x^2}) W^{*}_{\kappa_\sigma,\mu} W_{\kappa_\sigma,\mu}\bigg],~
\eea  
where $\x=\frac{k}{aH}$ and $\kappa_{\pm}= \mp i \xi$ and $\mu^2=\frac14-\frac{m^2_{W_R}}{H^2}$ which is presented in Fig. \ref{fig:cT}.

\begin{figure}[h!]
 \includegraphics[height=0.2\textheight]{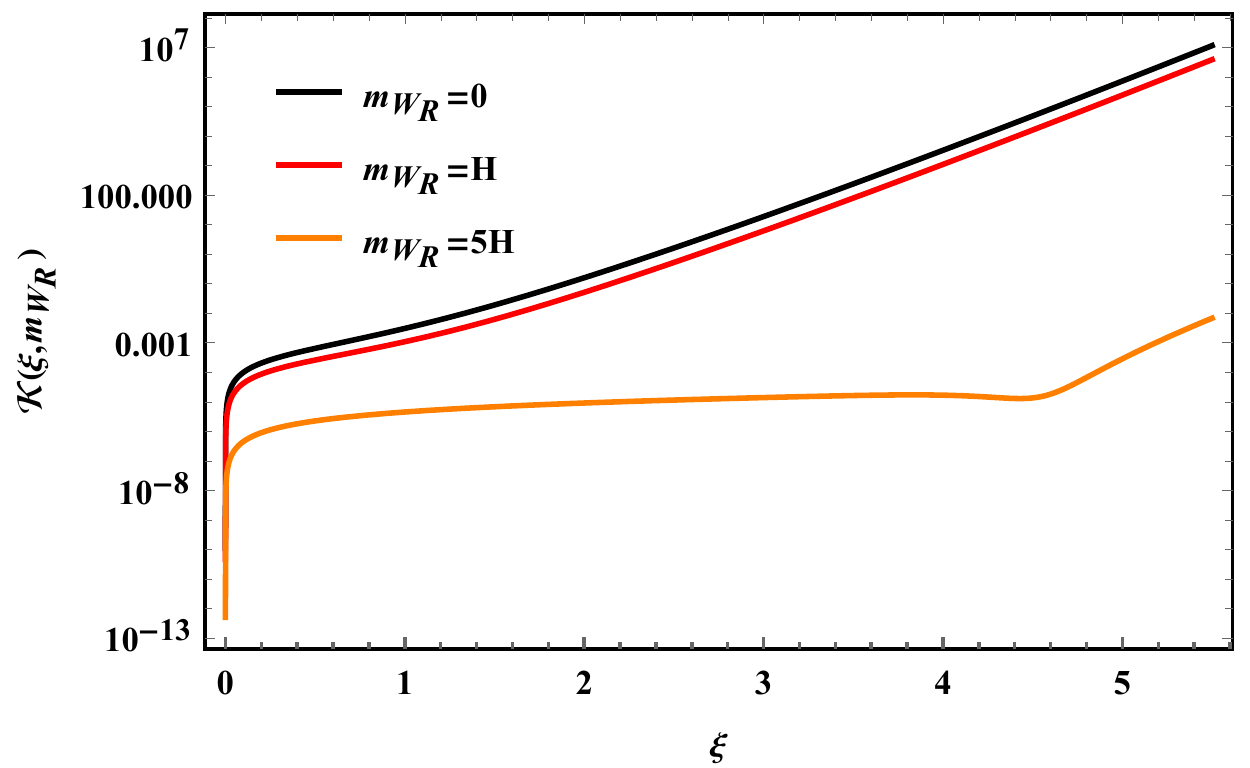}
 \includegraphics[height=0.2\textheight]{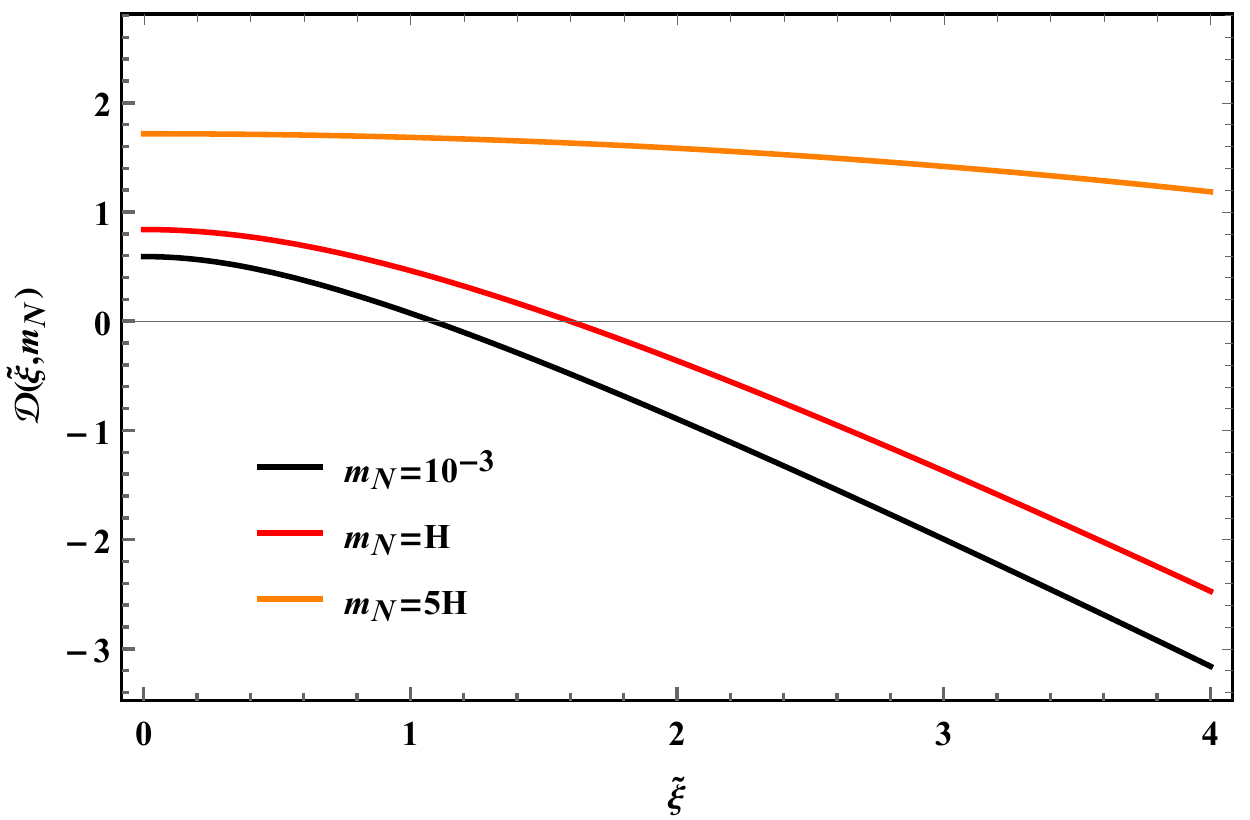}
\caption{The prefactors $\mathcal{K}(\xi,m_{W_R})$ and $\mathcal{D}(\tilde\xi,m_{N_i})$ in Eq.s \eqref{nB} and \eqref{nL}. The left panel shows $\mathcal{K}(\xi,m_{W_R})$ vs $\xi$ for different values of $m_{W_R}$ and the right panel presents $\mathcal{D}(\tilde\xi,m_{N_i})$ vs $\tilde\xi$ for different values of $m_{N_i}$.  }
\label{fig:cD} 
\end{figure}

The prefactor $\mathcal{K}(\xi,m_{W_R})$ in Eq.s \eqref{nB} and \eqref{nL} is the following momentum integral
\bea\label{cK--}
\mathcal{K}(\xi,m_{W_R}) \equiv  \frac{9}{4(2\pi)^4} \sum_{\sigma=\pm} \sigma e^{i\kappa_{\sigma}\pi} \int \x^3 d\ln\x W^{*}_{\kappa_{\sigma},\mu}(-2i\x)W_{\kappa_{\sigma},\mu}(-2i\x),
\eea
with the same $\mu$ and $\kappa_{\sigma}$ values as Eq. \eqref{cT-a}. $\mathcal{K}(\xi,m_{W_R})$ is shown in the left panel of Fig. \ref{fig:cD}.
 Moreover, the prefactor $\mathcal{D}(\tilde\xi,m_{N_i})$ in Eq.s \eqref{nL} is the contribution of axion in the production of massive sterile neutrinos (in type {\bf{II}} scenarios) in inflation. More precisely, apart from $W_R$, the massive fermions can get generated by the background axion field as well. However, the axion can not create chiral (massless) fermions. The reason is that a Peccei-Quinn type $U_{PQ}(1)$ \cite{Peccei:1977hhh} rotation of fermions as $
\Psi \rightarrow e^{-\frac{i\tilde\lambda}{f}\varphi} \Psi$, removes the fermion-axion interaction and simply transforms the fermion mass term as \cite{Weinberg:1996kr} $m_{\Psi} \rightarrow e^{\frac{2i\tilde\lambda}{f}\varphi} m_{\Psi}$. At the leading order, this effect can be captured as \cite{Maleknejad:2021nqi}
\bea\label{Int-n-m}  
\bar{n}_{N_i} \equiv \int d^3k \langle \bn_{iR}^{\dag} \bn_{iR}\rangle = - H^3 \bar{\xi} \sum_{i} \big(\frac{m_{N_i}}{H}\big)^2\mathcal{D}(\tilde\xi, m_{N_i}), 
\eea
where $\bar{n}_{N_i}$ is the the number density of massive RHNs generated by the background and the bar emphasises that, unlike chiral anomaly, it is a classical effect. This calculation is more involved and is done analytically by the author in \cite{Maleknejad:2021nqi}. Here we show $\mathcal{D}(\tilde\xi, m_{N_i})$ in the right panel of Fig. \ref{fig:cD}. For $\vert \tilde\xi\vert >1$, it has the following asymptotic forms 
\bea
 \mathcal{D}(\tilde\xi,m_{N_i}) \simeq \begin{cases}
\frac{ 2}{\pi} ~  \big[  \ln(\!\frac{m_{N_i}}{H}\!) - \psi^{^{\!(0)}}(1) + \frac12\big]  & ~~ (\frac{m_{N_i}}{H}\gg 1) ,\\
  -\frac{4}{3}   \lvert \tilde\xi \rvert  & ~~ (\frac{m_{N_i}}{H}\ll 1). 
\end{cases} 
\eea

\section{Spectator Effects in Leptogenesis Era}\label{Spectators}

Throughout the Early Universe, particles experience a whole cascade of interactions that eventually equilibrium in the Early Universe. Many of them can potentially redistribute the initial asymmetries to the spectator degrees of freedom.
The spectator effects in this scenario are studied in \cite{Maleknejad:2021nqi}.
Here we present a summary of these effects, i.e., sphaleron processes and lepton flavor effects. 

\begin{center}
{\bf{Flavor Effects:}}
\vspace{0.2mm}
\end{center}

Due to a CP-violating source, by the end of inflation we have a lepton quantum state $\lvert l_{\rm inf} \rangle$ as
\bea
\lvert l_{\rm inf} \rangle  \equiv  \sum_{\alpha=e,\mu,\tau} C^{\rm inf}_{\alpha} \lvert \alpha \rangle \where C^{\rm inf}_{\alpha} = \langle \alpha \vert l_{\rm inf} \rangle,
\eea
where $C^{\rm inf}_{\alpha}$ are specified by physics of inflation.
The composition of this primordial initial leptons and their CP conjugated anti-leptons are different. Moreover, the CP violating decays of the heavy sterile neutrinos can modify these initial states. The leptons produced in $\bN_i$ decays can be described in terms of quantum states denoted as $\lvert l_i \rangle$ that
can be decomposed in SM flavor space as
\bea
\lvert l_i \rangle \equiv \sum_{\alpha=e,\mu,\tau} C_{i\alpha} \lvert \alpha \rangle  \where C_{i\alpha} = \langle \alpha \vert l_{i} \rangle,
\eea
where $ C_{i\alpha}$ are coefficients given by the leptonic Yukawa matrix. Note that $\lvert l_i \rangle$s do not form an orthonormal bases, i.e. in general $\langle l_i \lvert l_{j\neq i} \rangle \neq 0$.
The processes has two stages of decay and wash-out, one for each of $\bN_3$ and $\bN_2$. Given the mass hierarchy considered in \eqref{mass-h-RHN}, at second stage with $T<10^{12}~GeV$, the $\tau$-lepton Yukawa interactions are thermalized. Hence the evolution can distinguish between $\tau$ flavor and $\tau^{\bot}=e+\mu$. That breaks the initial coherency between components parallel and orthogonal to $\tau$ which demands separate Boltzmann equations for each.

\begin{figure}[h!]
\centering
\includegraphics[height=0.25\textheight]{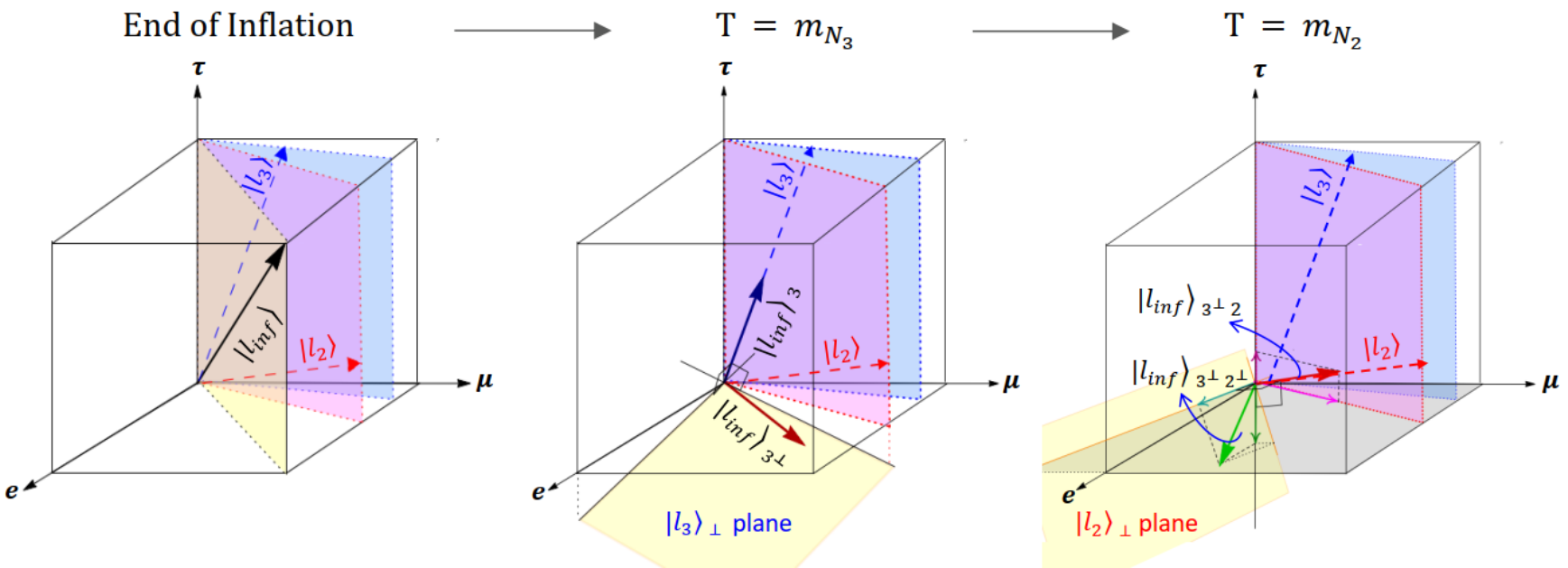}
\caption{ The geometrical illustration of flavor effects on washout processes of $\bN_3$ and $\bN_2$ in the SM flavor basis. The left panel shows the leptonic states at the end of inflation. The middle panel shows the lepton states at $T=m_{N_3}$ and the right panel presents the system at $T=m_{N_2}$. Note that total elimination of the pre-existing asymmetry requires highly fine-tuned relations between the flavored decay rates, hence on leptonic Yukawa couplings, and the flavor-space direction of the inflationary asymmetry. More precisely, one needs either i) $\vert l_{inf} \rangle$ coincides with one of $\vert l_{2} \rangle$ and $\vert l_{3} \rangle$, or ii)  $\vert l_{2} \rangle$ and $\vert l_{3} \rangle$ are perpendicular to each other which $\vert l_{inf} \rangle$ is in the plane of $\vert l_{2} \rangle-\vert l_{3} \rangle$.
}
\label{fig:flavor} 
\end{figure}

The decay of $\bN_{i}$ washes out the pre-existing asymmetry in the direction of heavy neutrino lepton flavor associated with its decay, i.e., $\vert l_i \rangle$. while leaves its perpendicular components, i.e. $\vert l_i \rangle_{\bot}$, unchanged. The geometry of this evolution in the SM flavor space is illustrated in Fig. \ref{fig:flavor}. After the second stage of RNH decays with $T<m_{N_2}$, the remnant asymmetry which remains unchanged by both washout effects is $\vert l_{\rm inf} \rangle_{3^{\bot}2^{\bot}}$. This flavor-vector consists of two incoherent components in $\tau$ and $\tau^{\bot}$ directions. Interestingly, eliminating the effect of this pre-existing asymmetry requires tightly fine-tuned relations between the flavored decay rates, hence on leptonic Yukawa couplings, as well as the flavor-space direction of the pre-existing lepton asymmetry. More precisely, one needs either i) $\vert l_{\rm inf} \rangle$ coincides with one of $\vert l_{2} \rangle$ and $\vert l_{3} \rangle$, or ii)  $\vert l_{2} \rangle$ and $\vert l_{3} \rangle$ are perpendicular to each other and $\vert l_{\rm inf} \rangle$ is in the plane of $\vert l_{2} \rangle-\vert l_{3} \rangle$. Given that $\lvert l_i \rangle$ and $\lvert l_{\rm inf} \rangle$ are specified by different physical parameters, these fine-tuning assumptions are extremely unnatural. In other words, the pre-existing asymmetry is memorized by cosmic evolution. In most of the parameter space, the remnant asymmetry, $n^{p,f}$, is significant, i.e.
\bea\label{pre-l}
n^{p,f}_{\mB-\uL}  \gtrsim \frac13 ~ n^{p,i}_{\mB-\uL},
\eea
where $n^{p,i}$ is the primordial value of asymmetry.

\begin{center}
{\bf{Sphaleron Effects:}}
\vspace{0.2mm}
\end{center}

Finally the $SU(2)_{L,R}$ sphalerons reshuffle the asymmetry of left-/right-handed leptons and quarks. In our setup the right-handed sphalerons are never in thermal equilibrium. Using the weak sphaleron effects and hypercharge constraint, we find that $\mB$, $\uL$, and $\mB-\uL$ are related as
\bea\label{n-sph}
n_{\mB} &=& c_{\rm sph} n_{\rm B-\uL},\\ \label{n-sph-2}
n_{\uL} &=&  (c_{\rm sph}-1) n_{\rm B-\uL},
\eea
where $c_{\rm sph}=\frac{28}{79}$ is the sphaleron conversion factor.  The combination of Eq.s \eqref{pre-l}-\eqref{n-sph-2} relates the final asymmetry to its primordial value as
\bea
n_{\mB}(a) &=& 0.12 ~ n^{\rm inf}_{\rm B-\uL} ~\big(\frac{a_{\rm inf}}{a}\big)^3,\\
n_{\uL}(a) &=&  -0.21 ~ n^{\rm inf}_{\rm B-\uL} ~\big(\frac{a_{\rm inf}}{a}\big)^3.
\eea

\section{Phenomenological Model of Reheating}\label{reheat}

\begin{figure}[h!]
\centering
\includegraphics[height=0.25\textheight]{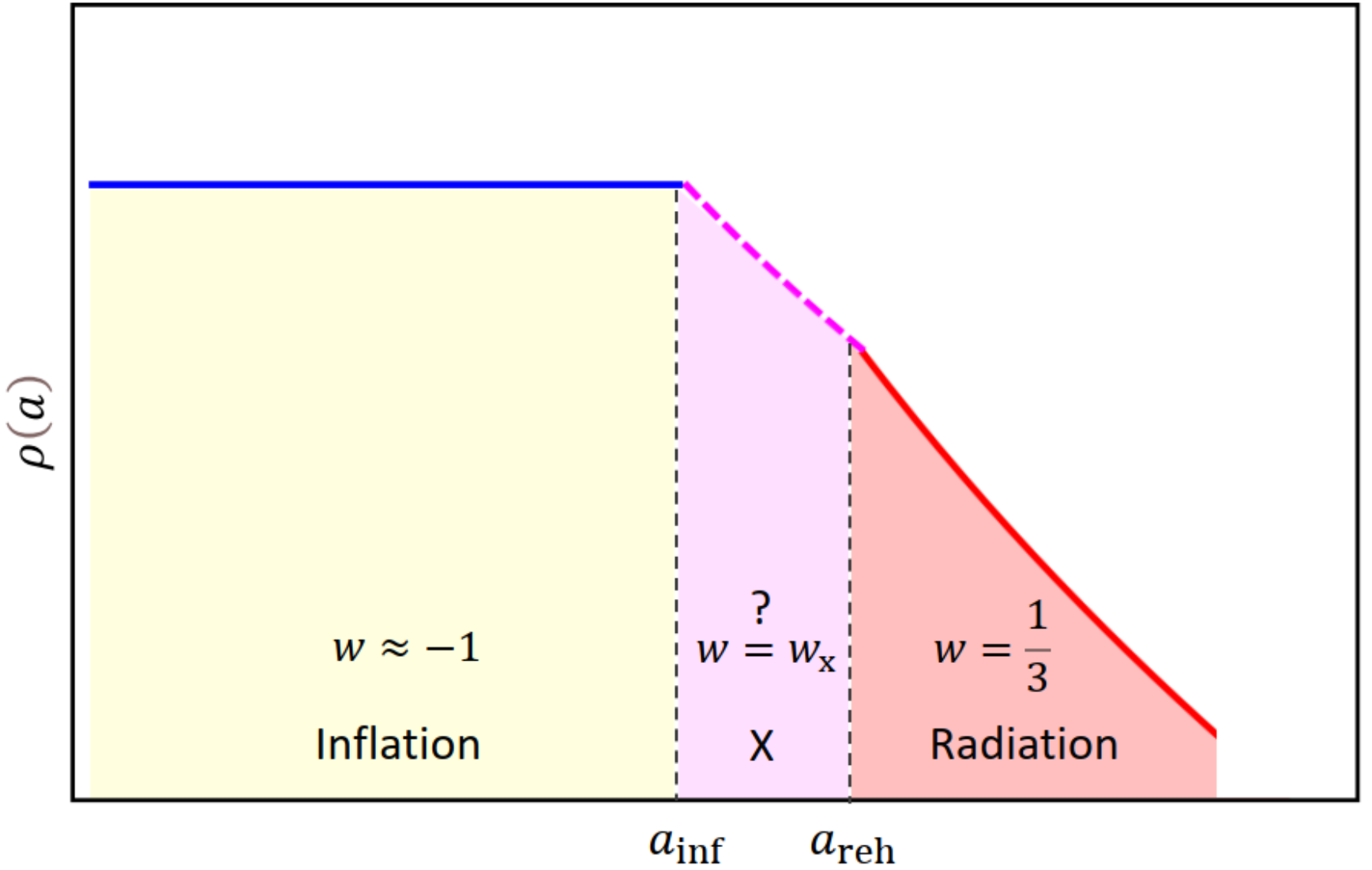}
\caption{The energy density of Universe vs scale factor. The dashed (pink) line which connects inflation to radiation era is a possible unknown intermediate phase with an average equation of state $w=w_X$.}
\label{omega-x} 
\end{figure}

Reheating starts at some point, $a_{\rm reh}$, after the end of inflation, $a_{\rm inf}$. Yet, the precise physics of reheating is not well understood. Here to quantify our analysis, we use a phenomenological model for reheating.
Depends on the details of the post-inflation physics, there may be an intermediate phase $X$ with the average equation of state $w_X$ which connects them (See Fig. \ref{omega-x}). In that case, the energy density of reheating is related to $\rho_{\rm inf}$ as
\bea
\rho_{\rm reh} \approx  \big(\frac{a_{\rm inf}}{a_{\rm reh}}\big)^{3(1+w_X)} ~ \rho_{\rm inf}.
\eea

We can write the above relation in the following phenomenological form
\bea
\rho_{\rm reh}=\varepsilon ~\rho_{\rm inf}~\big(\frac{a_{\rm inf}}{a_{\rm reh}}\big)^{4},
\eea
where $\varepsilon$ is the efficiency of the reheating process given as
\bea
\varepsilon \approx \big(\frac{a_{\rm inf}}{a_{\rm reh}}\big)^{3w_X-1}.
\eea

From the combination of Eq.s \eqref{C3} and \eqref{etaB}, we find
\bea\label{alpha-Delta-N}
\alpha_{\mB-\uL}^{\rm inf} \gtrsim 10^{-6} \mathcal{A}^{-\frac32}~e^{-\frac32(3w_X+1)\Delta N} \bigg(\gR \frac{\mpl}{m_{W_R}}\bigg)^4,
\eea
where $\mathcal{A}=\mathcal{O}(1)$, and $\Delta N$ is the number of e-folds between end of inflation to reheating, i.e.
\bea
\frac{a_{\rm inf}}{a_{\rm reh}} = e^{-\Delta N}.
\eea
The relation \eqref{alpha-Delta-N} provides the parameter space in which the freeze-out and freeze-in production of RHNs after inflation is negligible (condition C3), while the remnant of the primordial asymmetry has the right baryon to photon ratio.

\begin{figure}[htb]
\includegraphics[height=0.22\textheight]{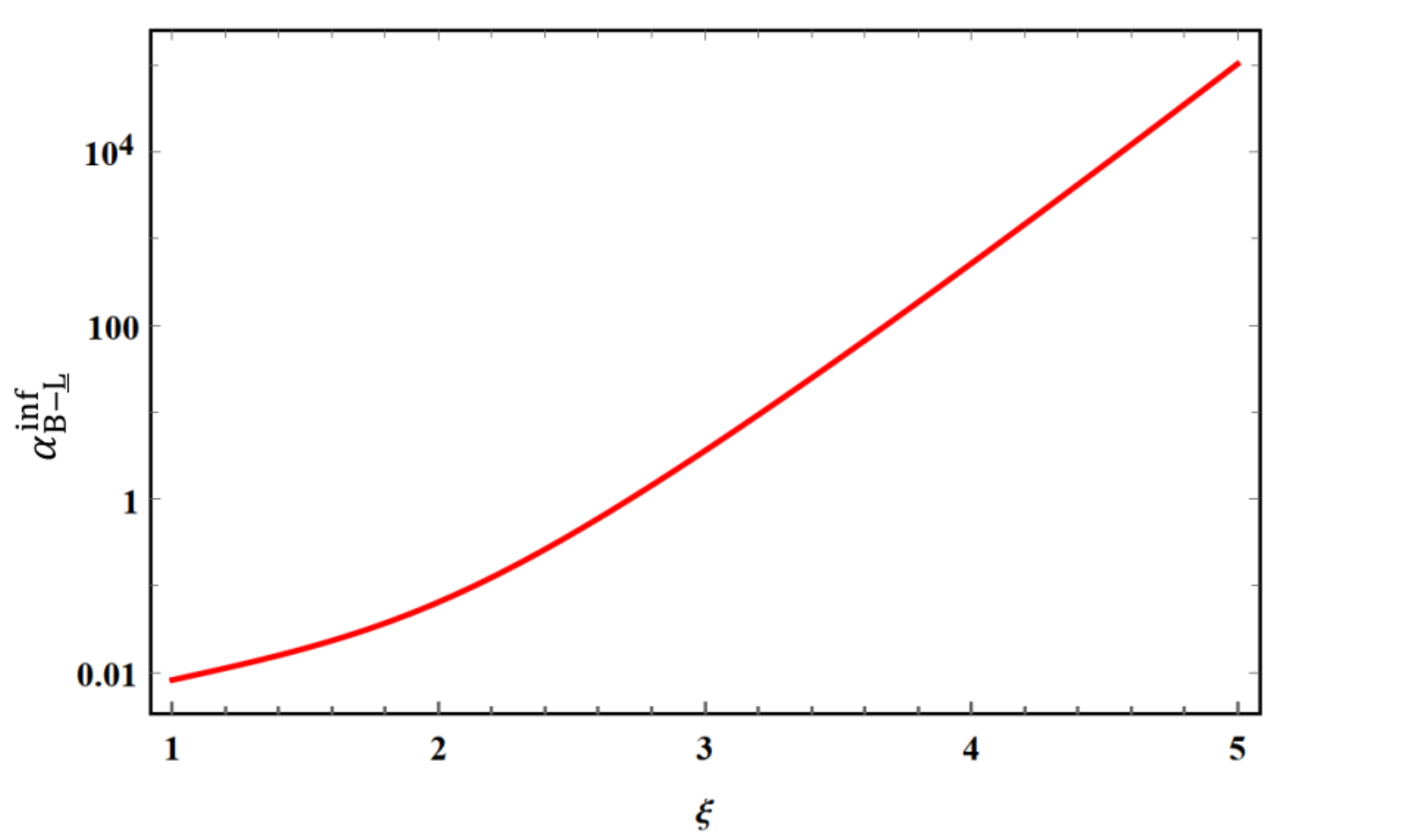}
\includegraphics[height=0.22\textheight]{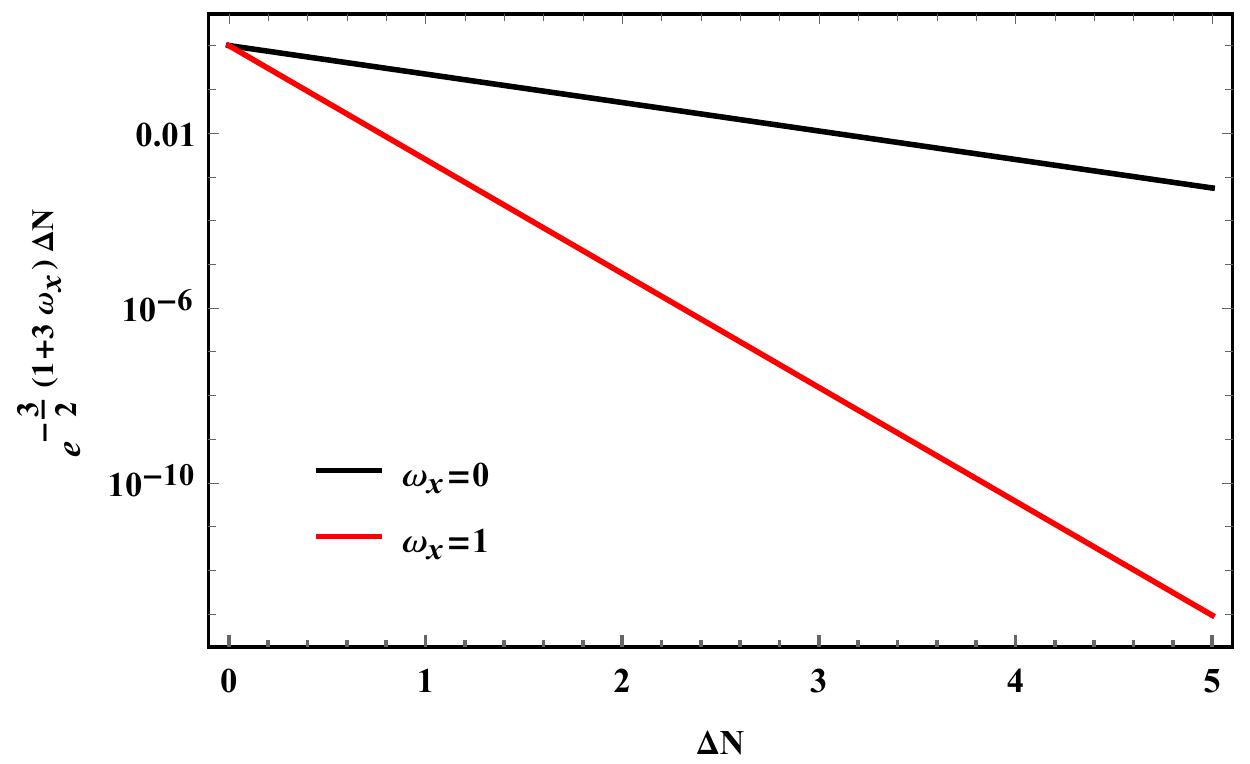}
\caption{ Left panel: the prefactor $\alpha_{\mB-\uL}^{\rm inf}$ in Eq. \eqref{alpha-def} vs $\xi$, with massless $W_R$ and $N_i$ in inflation. Right panel: the prefactor  $e^{-\frac32(3\omega_X+1)\Delta N}$ in Eq. \eqref{alpha-Delta-N} vs $\Delta N=\ln\big(\frac{a_{\rm reh}}{a_{\rm inf}}\big)$.}
\label{fig:alpha} 
\end{figure}

Two possible scenarios for the intermediate phase, i.e. $X$-era in Fig. \ref{omega-x}, are: S-i) inflation ends in a short period of matter domination with $\omega_X=0$ with reheating efficiency parameter as
\bea
\varepsilon \simeq \big(\frac{a_{\rm reh}}{a_{\rm inf}}\big) = e^{\Delta N} >1,
\eea
which implies 
\bea
\alpha_{\rm \mB-\uL}^{\rm inf} \gtrsim 10^{-6} ~e^{-\frac32\Delta N} \big(\gR\frac{\mpl}{m_{W_R}}\big)^{4},
\eea
or S-ii) inflation ends with domination of the kinetic term such that $\omega_X=1$ and $\varepsilon$ is
\bea
\varepsilon \simeq \big(\frac{a_{\rm inf}}{a_{\rm reh}}\big)^2 =e^{-2\Delta N} <1,
\eea
which gives
\bea
\alpha^{\rm inf}_{\rm \mB-\uL} \gtrsim 10^{-6} ~e^{-6\Delta N} \big(\gR\frac{\mpl}{m_{W_R}}\big)^{4}.
\eea
The necessary condition for C3 to hold is that $\Lambda_F \gtrsim \Lambda_{\rm inf}$. On the other hand, sufficient baryon asymmetry requires light $W_R$ in inflation (See Eq. \ref{Eta-1} and Fig. \ref{fig:cD}). The above demand $\Lambda_F\approx \Lambda_{\rm inf}$, i.e. the LR SSB should coincide with the end of inflation. More precisely, we need $H\mpl \approx \gR^{-2}m^2_{W_R}$ where $W_R$ is light in inflation and $m_{W_R}$ is the mass after inflation. It is indeed interesting that the LR symmetry breaking is related to a geometrical phase transition in cosmology, i.e. the end of exponential expansion of the Universe.

\end{document}